\documentclass[aps,prb,reprint,groupedaddress]{revtex4-2}
\usepackage{graphicx}
\usepackage{float}
\usepackage{subfigure}
\usepackage{amsmath}
\usepackage[colorlinks=true, allcolors=blue]{hyperref}
\usepackage{tikz}
\usetikzlibrary{shapes.geometric, arrows}
\tikzstyle{startstop} = [rectangle, rounded corners, minimum width=1cm, minimum height=0.5cm,text centered, draw=black,line width=0.35mm]
\tikzstyle{process} = [rectangle, minimum width=3cm, minimum height=0.5cm, text centered, draw=black,line width=0.35mm]
\tikzstyle{decision} = [diamond, aspect=3, minimum width=1cm, minimum height=0.5cm, text centered, draw=black,line width=0.35mm]
\tikzstyle{arrow} = [thick,->,>=stealth,line width=0.35mm]

\begin{document}
\preprint{}

\title{Nonequilibrium mean-field approach for quantum transport with off-diagonal disorder}


\author{Rongjie Cui\textsuperscript{1}}
\author{Zelei Zhang\textsuperscript{1}}
\author{Qi Wei\textsuperscript{2}}
\author{Yu Zhang\textsuperscript{3}}
\author{Youqi Ke\textsuperscript{1}}\email[]{keyq@shanghaitech.edu.cn}
\affiliation{\textsuperscript{1}School of Physical Science and Technology, ShanghaiTech University, Shanghai 201210, China}
\affiliation{\textsuperscript{2}The Guangdong Institute of Intelligence Science and Technology, Zhuhai 519031, China}
\affiliation{\textsuperscript{3}National Supercomputing Center in Shenzhen, Shenzhen 518055, China}


\date{\today}

\begin{abstract}
For the nanoscale structures, disorder scattering plays a vital role in the carriers' transport, including electrons and high-frequency phonons. The capability for effectively treating the disorders, including both diagonal and off-diagonal disorders, is indispensable for quantum transport simulation of realistic device materials. 
In this work, we report a self-consistent nonequilibrium mean-field quantum transport approach, by combining the auxiliary coherent potential approximation (ACPA) and non-equilibrium Green's function method, for calculating the phonon transport through disordered material structures with the force-constant disorders (including the Anderson-type disorder). The nonequilibrium vertex correction (NVC) is derived in an extended local degree of freedom to account for both the multiple disorder scattering by force-constant disorder and the nonequilibrium quantum statistics. We have tested ACPA-NVC method with the fluctuation-dissipation theorem at the equilibrium and obtained very good agreement with supercell calculations for the phonon transmission. To demonstrate the applicability, we apply ACPA-NVC to calculate the thermal conductance for the disordered Ni/Pt interface, and important effects of force-constant disorder are revealed. ACPA-NVC  method  provides an effective quantum transport approach for simulating disordered nanoscale devices, and the generalization to simulate disordered nanoelectronic device is straightforward.
\end{abstract}


\maketitle

\section{INTRODUCTION}
Disordered impurities or defects inevitably exist in various device materials, e.g.,nano-transistors, photodiodes, thermoelectric devices, phase-change memory,\cite{apply},  and the transport properties can be significantly modulated or altered by the random impurities/defects inside the devices.\cite{Transport1,Transport2,Transport3} It is therefore imperative to completely understand the effects of various disorders on carrier transport from both the experimental measurement and theoretical simulations. Because of the breakdown of translational symmetry for the presence of disorders, meaningful simulations are required to calculate the average of physical properties over disorder configurations, presenting important challenges for simulations. Over the past decades, considerable efforts have been spent to develop effective methods to derive the disorder average to enable the simulation of disorders. The direct method utilizes the supercell(SC) with the randomly sampled or special-quasirandom configurations\cite{SQS}, while the large computational cost and  finite supercell size limit its applicability. The important self-consistent dynamical-type mean field approach, named the coherent potential approximation (CPA)\cite{CPA1,CPA2}, have been developed to simulate the disorders. CPA self-consistently construct an translational invariant effective medium with correct analytical properties.\cite{Leath1974,b.velicky1968} CPA is well-known for its easy implementation and high efficiency with important accuracy in physical properties in single-site approximation(SSA).\cite{b.velicky1968,SSA} Since the first proposal for both the disordered vibrational and electronic systems in 1960s,\cite{CPA1,CPA2} great extensions to CPA have been made to improve the accuracy and expand the applicabilities. To account for the nonlocal correlation effects of disorder scattering, the important cluster extensions, including molecular CPA(MCPA) and dynamical cluster approximaiton(DCA),\cite{MCPA1,MCPA2,DCA1,DCA2,ZhaiJX} and diagrammatic techniques\cite{Leath1974} have been developed and found important success for many applications. In combination with the nonequilibrium Green's function (NEGF)\cite{NEGF1,NEGF2} method, CPA and its cluster extensions have been developed to simulate the quantum transport through disordered nanoelectronics.\cite{Zhangyu,NVC1,HuCY} The full counting statistics CPA has been developed to calculate the disorder average of high-order cumulants of current.\cite{FullC1,FullC2} Moreover, the combination with the density functional theory (DFT) great extend the capability of CPA and its various extensions for first-principles simulation of realistic disordered materials and devices,including notably the Korringa-Kohn-Rostocker and Muffin-Tin Orbital methods.\cite{KKR-CPA,FirstP5,Zhangyu,ZhangQY,YanJW}

However, CPA and its various cluster extensions, as mean-field approaches, are essentially based on the embedding framework\cite{embedding1,embedding2}, providing a local approximation to do disorder average. However, the embedding framework in the physical space is suitable for handling the disorders on the limited local degree of freedom, namely the diagonal disorder\cite{diagonal3,diagonal4,diagonal5,diagonal6}, rather than the off-diagonal disorder which is nonlocal and is also generally existing (e.g. the random hopping in electronic systems and force-constant disorder (FCD) in vibrational systems). The self-consistent mean-field approach for treating the general off-diagonal disorders is thus a long-standing issue for the theoretical study of disordered systems. Without properly treating off-diagonal disorders, mean-field approaches can fail to reproduce the correct properties of disordered alloys.\cite{FCD1,FCD2,FCD3} Great theoretical efforts have been made to extend the mean-field approach to off-diagonal disorders, including the Blackman-Esterling-Berk(BEB) transformation based CPA \cite{BEB} and itinerant CPA (ICPA)\cite{ICPA}. BEB transformation changes the off-diagonal disorder to be nonrandom in the extended Hamiltonian in an augmented space, but limits the applicability to uncorrelated diagonal and off-diagonal disorders, precluding the disordered lattice vibrational problem. ICPA is also based on the augmented space and provide a general way to treat the off-diagonal disorder, but features higher complexity for implementation and computation compared to conventional CPA method. Both BEB-CPA and ICPA algorithms possess important difficulty to include the Anderson-type off-diagonal disorder with a distribution, due to the significant increase of the augmented space. 

Recently,  a method, called the auxiliary coherent potential approximation (ACPA)\cite{ACPA1,ACPA2,ACPA3,ACPA4,ZhaiJX,Zhangzl} is proposed by Ke and coworkers, to treat both the general diagonal and off-diagonal disorders (including force-constant and hopping disorders). The ACPA
is based on introducing an decomposition
that transforms the off-diagonal disorder into a diagonal-like disorder of an auxiliary system, and
then the self-consistent CPA algorithm with an extended local degree of freedom can be applied to obtain
the auxiliary coherent medium to give averaged physical
properties. The ACPA features advantages of easy implementation and high computational efficiency.\cite{ACPA1,ACPA2} The applicability of ACPA has been well demonstrated for the disordered lattice vibration of alloys with force-constant disorder (FCD). It's been proved that ACPA
presents very good agreement with ICPA and experimental
measurements in calculating the phonon dispersions of different fcc alloys\cite{ACPA2}. Moreover, ACPA can be generalized to effectively treat the Anderson-type force-constant disorder (FCD) with large distribution due to environmental disorder.\cite{ACPA4} By introducing the cluster coupling space, the cluster extension of ACPA, namely auxiliary dynamical cluster theory, has been developed to account for the nonlocal correlation effects of off-diagonal disorder.\cite{ZhaiJX} Extending ACPA to electronic system has provided important applications for dynamical mean-field simulation of the Anderson-Hubbard model with off-diagonal disorder.\cite{Zhangzl} In this work, we report the extension of ACPA method in combination with nonequilibium Green's function method(NEGF) to enable the quantum transport simulation of the realistic nano-structures with atomic disorders. 

The rest of the paper is organized as follows:  Sec.\ref{method} presents the formulation and implementation of the ACPA, and the nonequilibrium vertex correction (NVC) for quantum transport simulation of a disordered two-probe structure. In Sec.\ref{results}, we validate the ACPA-NVC method for phonon transport simualtion with force-constant disorders (including the Anderson type), and apply it to calculate the phonon transmission and the phonon conductance for the disordered Ni/Pt interface. At last, we conclude our work in section \ref{conclusion}.

\section{FORMALISM}\label{method}
\subsection{ Force-constant decomposition and auxiliary coupling space}

To deal with the disordered lattice vibration, we first introduce an separable model for the force constant disorder (FCD) in the following format
\begin{equation}\label{FCDmodel}
k^{QQ\prime,n}_{ij}=x^Q_{i}S_{ij}^nx^{Q\prime}_{j}+\lambda_{ij}^n,
\end{equation}
where $x^{Q/Q\prime}_{i/j}$ is dependent on the atomic occupant $Q/Q\prime$ on the
site $R_i/R_j$ with the concentration denoted as $c_i^{Q/Q'}$, the quantities $S_{ij}^n$and $\lambda_{ij}^n$ are independent of the
atomic occupations on sites $R_i$ and $R_j$ and the superscript $n$ accounts for the force-constant distribution in $k_{ij}^{QQ\prime}$ with the probability $p^n$ and $\sum_np^n=1$$(n=1,2,...,N)$, describing the Anderson-type FCD (AFCD). \cite{ACPA4} It has been demonstrated that Eq.(\ref{FCDmodel}) can cover a large range of parameter space for the FCD\cite{ACPA2,ACPA3,ACPA4}.

\begin{figure}[htbp]
			\centering
  \includegraphics[width=0.5\textwidth]{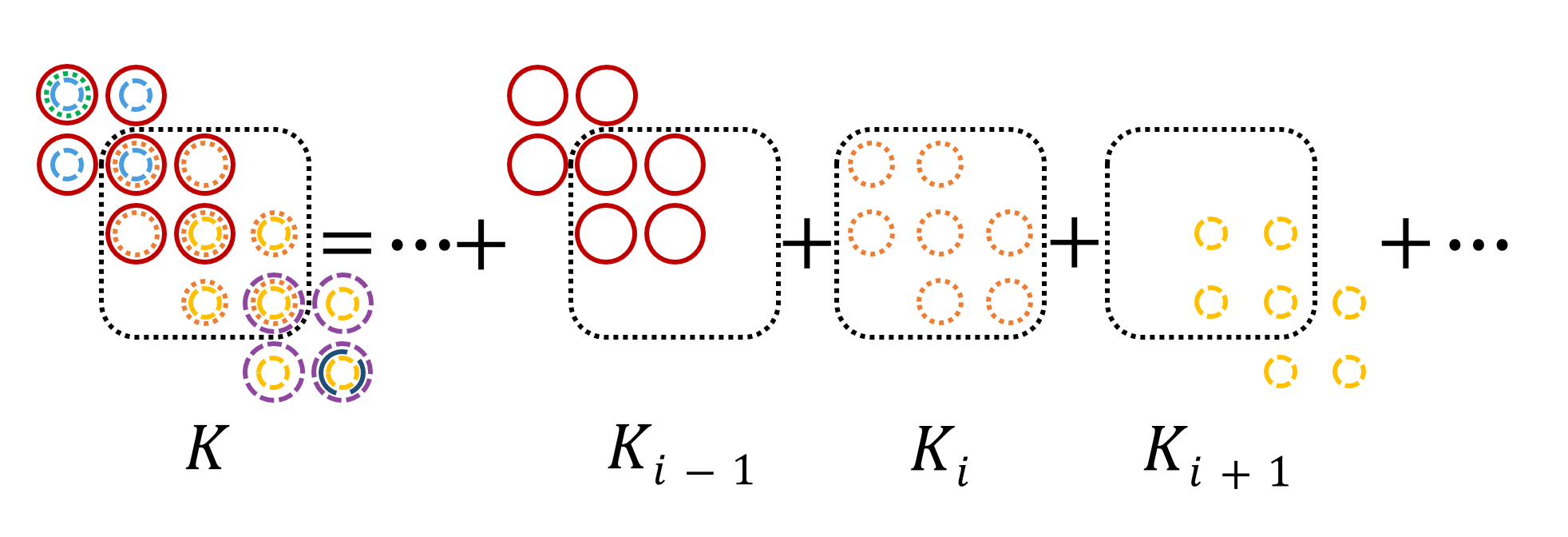}
    \caption{Schematic illustration of the summation for the auxoliary quantities. The circles with different color represent the nonzero matrix elements of $K$ and $K_i$ matrixes}
\label{sum}
\end{figure}

As a result, we can rewrite the dynamic matrix $D$ as
\begin{equation}
D=XK,
\end{equation}
where $X$ is a diagonal matrix with $X_{ii}=x_{i}$ and $K$ represents an auxiliary system. Here, the $K$ matrix can be rewriten as a
 sum of single-site quantities
\begin{equation}\label{KiSS}
K=\sum_{i}K_{i},
\end{equation}
where the singel-site quantity ${K}_i$ depends on the atomic species at the site $i$. Thus,with the Eqs.(\ref{FCDmodel})-(\ref{KiSS}), we transform
the FCD in physical space into a diagonal-like disorder in an auxiliray system $K$. The dimension of ${K}_i$ matrix is $d(Z+1)$ where $d$ is the single-site degree of freedom, and $Z$ is the number of neighbors with nonzero force-constant with the site $i$. Taking 1D atomic chains with nearest-neighbor FCD as an example, we can schematically show the summation in Eq.(\ref{KiSS}) in Fig.\ref{sum}.

To describe the single-site quantities in auxiliary system with the extended local dimension of $d(Z+1)$ , we introduce the coupling space $\mathcal{C}$ for each lattice site coupling to $Z$ neighboring sites.\cite{ZhaiJX} In the coupling space, we can define ${K}_{i}$ as
\begin{equation}
\begin{aligned}
{K}_{i}^{\mathcal{C}}& =\sum_{j\ne0}S_{j i} x_{i}\mid{\vec{T}}_j\rangle \langle{\vec{T}}_j\mid
-\sum_{j\ne0}S_{j i} x_{i}\mid{\vec{T}}_j\rangle \langle{\vec{T}}_0\mid \\
&-\sum_{j\ne0}\frac{\lambda_{i j}}{x_{i}}\mid{\vec{T}}_0\rangle \langle{\vec{T}}_j\mid
+\sum_{j}\frac{\lambda_{j i}}{x_{i}}\mid{\vec{T}}_0\rangle \langle{\vec{T}}_0\mid,
\end{aligned}
\end{equation}
where  $\vec{T}_j$ is the translational vector that relates the neighboring sites $R_j$ to the site $R_i$,e.g.$\vec{R}_j=\vec{R}_i+\vec{T}_j$, and the bases $|\vec{T}_j\rangle$ encompass the coupling information of the site $R_i$ along the $\vec{T}_j$ direction. It should be mentioned that, as an important feature of the above force-constant decomposition model, the force-constnat sum-rule is always conserved for obeying the translational symmetry of physics.\cite{ACPA3} With this definition, we can rewrite Eq.(\ref{KiSS}) as
\begin{equation}
K_{mn}=\sum_{i}K^{\mathcal{C}}_{i,T_jT_k},\label{auxiSum}
\end{equation}
where $K^{\mathcal{C}}_{i,T_jT_k}=\langle{\vec{T}}_j\mid K_i^{\mathcal{C}}\mid{\vec{T}}_k\rangle$, $\vec{R}_m=\vec{R}_i+\vec{T}_j$ and $\vec{R}_n=\vec{R}_i+\vec{T}_k$. In Eq.\ref{auxiSum}, to obtain the matrix element $K_{mn}$ in the matrix $K$, we sum all the terms in the coupling space for all sites $R_i$ that satisfy the relations $\vec{R}_m=\vec{R}_i+\vec{T}_j$ and $\vec{R}_n=\vec{R}_i+\vec{T}_k$ . As an example of a one-dimensional system with nearest-neighbour FCD, as shown in Fig.\ref{sum}, $K_{00}$ can be calculated by $K_{00}=K^{\mathcal{C}}_{0,T_0T_0}+K^{\mathcal{C}}_{-1,T_1T_1}+K^{\mathcal{C}}_{1,T_{-1}T_{-1}}$ . Hereafter all the quantities with the superscript $\mathcal{C}$ indicate that they are defined in the coupling space and satisfy similar summation relations. For more details about the coupling space and the extended local degree of freedom, please refer to the Ref.\onlinecite{ZhaiJX}.

\subsection{Auxiliary Green's function and phonon transport }

\begin{figure}[htbp]
    \centering
    \includegraphics[width=0.5\textwidth]{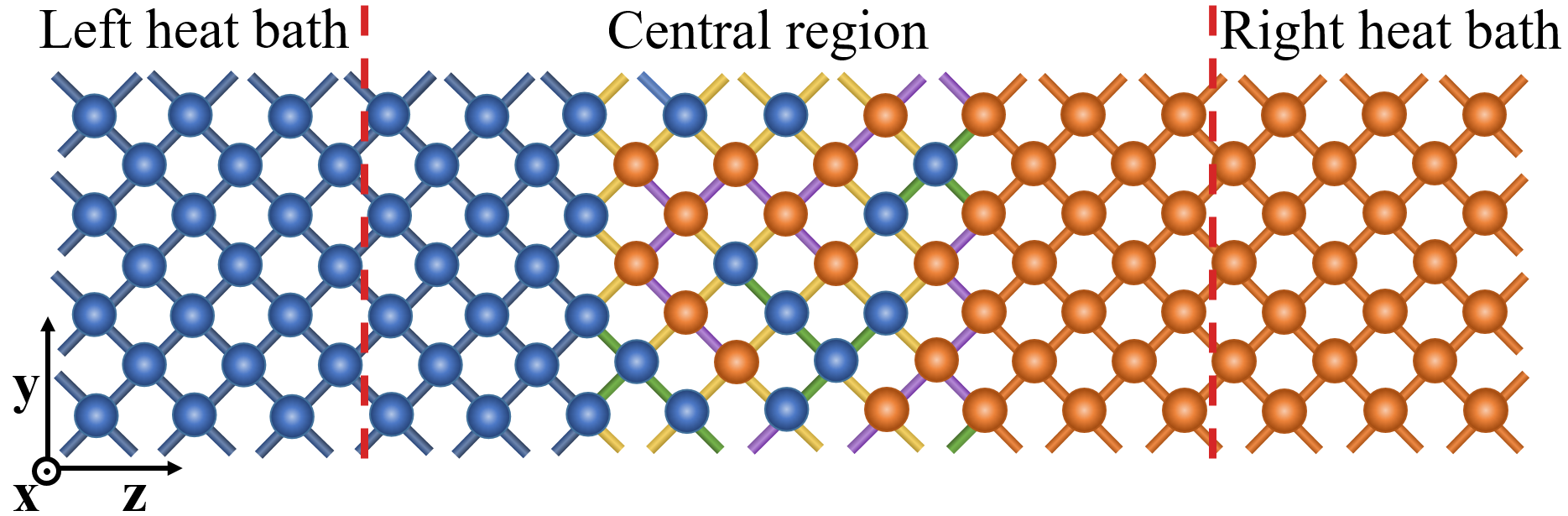}
    \caption{Schematic illustration of the disordered two-probe device structure $Ni|Ni_{x}Pt_{1-x}|Pt$ . The spheres and bonds with different colors in the central region denote the respective mass disorder and force constant disorder due to the random distribution of Ni and Pt atoms.}
\label{IS}
\end{figure}

We consider a two-probe device structure containing a disordered central region sandwiched by the left and right heat bathes, as shown in Fig.\ref{IS}. In the central region, the atomic disorders contain both the mass disorder and FCD. It is known, for a specific atomic configuration, the Green's function of lattice vibration can be written as, for the two-probe device,
\begin{equation}
G=\frac{1}{m\omega^2-D-\Sigma_L-\Sigma_R},
\end{equation}
where m is the atomic mass, $\omega$ is the frequency of phonon, $\Sigma_L$,$\Sigma_R$ are the self-energies of left and right bathes, respectively.

With the force-constant decomposition we can rewrite the Green's function 
as
\begin{equation}
G=gX^{-1},
\end{equation}
with an auxiliary GF defined as
\begin{equation}
g=\frac{1}{X^{-1}m\omega^2-K-\tilde{\Sigma}_L-\tilde{\Sigma}_R},
\end{equation}
where $\tilde{\Sigma}_L=X^{-1}\Sigma_L$and $\tilde{\Sigma}_R=X^{-1}\Sigma_R$ are the auxiliary 
self-energies which is independent of disorders. The corresponding auxiliary Hamiltonian of the central device can be defined as 
\begin{equation}\label{P1}
P=X^{-1}m\omega^2-K.
\end{equation}
Similar to $K$, $P$ can be expressed as a sum of single-site quantity $P_i^{\mathcal{C}}$, namely
\begin{equation}\label{P2}
P=\sum_{i}P_{i}^{\mathcal{C}},
\end{equation}
and 
\begin{equation}\label{P3}
P_{i}^{\mathcal{C}}=x_{i}^{-1}m_{i}\omega^2-K_{i}^{\mathcal{C}},
\end{equation}
in which $P_{i}^{\mathcal{C}}$ is only dependent on the atomic occupation of the site $R_i$. As a result, with Eqs.(\ref{P1})-(\ref{P3}), the general disorder problem in lattice vibration is transformed to an auxiliary diagonal-like disorder problem, enabling the implementation of conventional CPA.\cite{CPA1,CPA2}

For the phonon transport, from the  Landauer's formula\cite{Land,Caroli}, the thermal conductance is given as
\begin{equation}
\sigma(T)=\frac{1}{2 \pi S} \int_0^{\infty} \hbar \omega \frac{\partial {f}(\omega, T)}{\partial T}  \Xi(\omega) d \omega 
\label{sigmaT}
\end{equation}
where $f(\omega, T)$ is the Bose-Einstein distribution function, $S$ is the cross-section area perpendicular to the transport direction,  and $\Xi(\omega)$ is the frequency-dependent transmission across the device.
With the non-equilibrium Green's Function (NEGF) method, the phonon transmission is given by the Caroli formula\cite{Caroli}
\begin{equation}
 \Xi(\omega)=Tr[G^R\Gamma_LG^A\Gamma_R ]
\end{equation}
where $G^{R/A}$ is the retarded/advanced physical Green’s function of
the central region, the linewidth functions $\Gamma_{L/R}$
describe the couplings of the central region to the left/right leads. By using the relation $G^{R/A}=g^{R/A}X^{-1}$, the averaged transmission function can be transformed to 
\begin{equation}
 \Xi(\omega)=Tr[{g}^R\tilde{\Gamma}_L{g}^A\tilde{\Gamma}_R]
\label{trans}
\end{equation}
where  $\tilde{\Gamma}_{R/L}=i[\tilde{\Sigma}_{R/L}^R-\tilde{\Sigma}_{R/L}^A]$ are the auxiliary linewidth functions which are independent of disorder. For a disordered device, physical quantities require the configurational average to be meaningful. 

\subsection{Auxiliary coherent potential approximation}
The central idea of CPA is to construct a coherent effective medium whose GF is equal
to the disorder-averaged GF of the system. To continue, one can introduce an effective auxiliary medium 
Hamiltonian 
\begin{equation}
\mathcal{P}=\sum_i\mathcal{P}_{i}^{\mathcal{C}},
\end{equation}
with the corresponding averaged auxiliary GF given by
\begin{equation}
 \bar{g}=\mathcal{P}^{-1}.
\end{equation}
Then for a specific configuration, the Dyson equation for auxiliary GF can be written as 
\begin{equation}
g=\bar{g}+\bar{g}Vg
\end{equation}
where $V=\sum_i(\mathcal{P}_{i}^{\mathcal{C}}-P_{i}^{\mathcal{C}})$ accounts for all the random fluctuations. By defining the T-matrix $T=V(1-\bar{g}V)^{-1}$, which contains all the information of multiple disorder scattering, the auxiliary Green's function can be rewritten as 
\begin{equation}\label{ga}
g=\bar{g}+\bar{g}T\bar{g}
\end{equation}
for which the disorder average gives $\langle g\rangle=\bar{g}+\bar{g}\langle T\rangle\bar{g}$.
By applying the relation $\langle g\rangle=\bar{g}$, we can obtain  the CPA condition  $\langle T\rangle=0$. Here, the solution of CPA equation $\langle T\rangle=0$ to obtain the effective $\mathcal{P}$ requires appropriate approximation. To do so, one can rewritten the T-matrix as 
\begin{equation}
T=\sum_i t_{i}^{\mathcal{C}}+\sum_{i \neq j} t_{i}^{\mathcal{C}} \bar{g} t_{j}^{\mathcal{C}}+\sum_{i \neq j, j \neq k} t_{i}^{\mathcal{C}} \bar{g} t_{j}^{\mathcal{C}} \bar{g}t_{k}^{\mathcal{C}}+\ldots 
\end{equation}
where $t_{i}^{\mathcal{C}}$ is the single-site scattering matrix in the coupling space and 
\begin{equation}
t_{i}^{\mathcal{C}}=({\mathcal{P}}_{i}^{\mathcal{C}}-{P}_{i}^{\mathcal{C}})[1-\bar{g}({\mathcal{P}}_{i}^{\mathcal{C}}-{P}_{i}^{\mathcal{C}})]^{-1}.
\label{ti}
\end{equation}

By introducing the SSA that decouples the correlation of scattering ${t}_{i}^{\mathcal{C}}$ of different sites, the CPA
condition is reduced to a single-site equation
\begin{equation}
\langle t_{i}^{\mathcal{C}}\rangle=0,
\end{equation}
with which the single-site quantity $\mathcal{P}_{i}^{\mathcal{C}}$ can be solved self-consistently, forming a self-consistent mean-field treatment of the disorders. In actual calculation, in order to make the calculation process stable, the locator-CPA\cite{inter} is often
applied.

Due to the restored periodicity in the x-y plane of the two-probe device after disorder averaging, we can apply two-dimensional (2D)
lattice Fourier transformation to the effective medium Hamiltonian $\mathcal{P}$ so that
\begin{equation}
\mathcal{P}_{\vec{B}_m,\vec{B}_n}( \vec{k}_{\|})=\sum_{\vec{T}_{\|}} e^{i \vec{k}_{\|} \cdot\vec{T}_{\|}} \mathcal{P}_{\vec{B}_m, \vec{B}_n+\vec{T}_{\|}}
\label{FT}
\end{equation}
where the vectors $\vec{B}_m,\vec{B}_n$ denote the atomic sites within the primitive 2D cell, $\vec{k}_{\|}$ is the wave vector in the 2D Brillouin zone, and $\vec{T}_{\|}$ is the two-dimensional translational vector and the dimension of $\mathcal{P}(\omega, \vec{k}_{\|})$ matrix is $Nd$ ($N$ is the total number of sites within the primitive cell of central region).

Then the averaged auxiliary GF for each $\vec{k}_{\|}$ is
\begin{equation}
\bar{g}(\vec{k}_{\|})=[\mathcal{P}(\vec{k}_{\|})-\tilde{\Sigma}(\vec{k}_{\|})]^{-1}
\label{gk}
\end{equation}

Within the CPA, to form a set of self-consistent equations to determine the effective medium Hamiltonian $\mathcal{P}$, a quantity called coherent interactor $\Delta_{i}^{\mathcal{C}}$ is introduced to describe the influence of the surrounding coherent medium to the site $i$. Then we obtain
\begin{equation}
\Delta_{i}^{\mathcal{C}}=\mathcal{{P}}_{i}^{\mathcal{C}}-[\bar{g}]_{i}^{\mathcal{C},-1}.
\label{delta}
\end{equation}
where $[\bar{g}]_{i}^{\mathcal{C}}$ is a $d(Z+1)\times d(Z+1)$ matrix block defined in the coupling space $\mathcal{C}$ associating with the site $i$, in which the element can be obatined by 
\begin{equation}
[\bar{g}]^{\mathcal{C}}_{i,\vec{B}_m+\vec{T}_{\|}, \vec{B}_n+\vec{T}_{\|}^{\prime}}=\frac{1}{\mathcal{V}} \int d \vec{k} \bar{g}_{\vec{B}_m, \vec{B}_n}(\vec{k}_{\|}) e^{-i \vec{k}_{\|}\cdot(\vec{T}_{\|}^{\prime}-\vec{T}_{\|})}
\label{gii}
\end{equation}
where $\vec{B}_m+\vec{T}_{\|}, \vec{B}_n+\vec{T}_{\|}^{\prime}$ denotes the lattice sites including the $i$ site and its neighbors with nonzero force constant. With the interactor, the conditionally averaged auxiliary GF, which corresponds to the system with the fixed ${P}_{i}^{\mathcal{C},Q,n}$ on the site $i$, can be given as
\begin{equation}
[\bar{g}]_{i}^{\mathcal{C},Q,n}=({P}_{i}^{\mathcal{C},Q,n}-\Delta_{i}^{\mathcal{C}})^{-1}
\end{equation}
and it's known that, with these expressions,
the new single-site quantities $\mathcal{P}_{i}^{\mathcal{C}}$ can be calculated by 
\begin{equation}
{\mathcal{P}}_{i}^{\mathcal{C}}=\left[\sum_{Q,n} c_{i}^{Q}p_i^n [\bar{g}]_i^{\mathcal{C},Q,n}\right]^{-1}+\Delta_{i}^{\mathcal{C}}
\label{Pi}
\end{equation}
The above Eqs.(\ref{FT})-(\ref{Pi}) form a complete CPA self-consistent loop to solve for $\mathcal{P}_{i}^{\mathcal{C}}$. To accelerate the convergence process, we applied the Andersen mixing method to mix the $\mathcal{P}_{i}^{\mathcal{C}}$ for next iteration. When the convergence is reached, we have the relation 
\begin{equation}
[\bar{g}]_{i}^{\mathcal{C}}=\sum_{Q,n}c_{i}^{Q}p_i^{n}[\bar{g}]_{i}^{\mathcal{C},Q,n}.
\end{equation}

\subsection{Nonequilibrium vertex correction theory}
To treat the disorder average of the product of two auxiliary Green's functions in the NEGF and transmission function of
 Eq.(\ref{trans}), we introduce the nonequilibrium vertex correction(NVC) \cite{NVC1,NEGF2} for treating the nonequilibrium off-diagonal disorder problem. With Eq.(\ref{ga}) and $\langle T^{R/A}\rangle=0$, we have
\begin{equation}
\langle{g}^R C {g}^A\rangle=\bar{g}^R C \bar{g}^A+\bar{g}^R\Omega\bar{g}^A
\label{GRCGA}
\end{equation}
where C is a contanst quantity related to perfect electrode and $\Omega=\langle T^R\bar{g}^R C \bar{g}^AT^A\rangle$ is the NVC accounting for the multiple disorder scattering at the nonequilibrium and can be rewritten as
\begin{equation}
\Omega=\sum_{i,j}\langle T^R_i\bar{g}^R C \bar{g}^AT^A_j\rangle
\label{Om}
\end{equation}
where 
\begin{equation}
T^{R/A}_i=t_i^{\mathcal{C},R/A}+\sum_{j\ne i} T^{R/A}_j\bar{g}^{R/A}t^{\mathcal{C},R/A}_i
\label{Ti}
\end{equation}
In the SSA, for $i\ne j$, we have 
\begin{equation}
\sum_{i\ne j}\langle T^R_i\bar{g}^R C \bar{g}^AT^A_j\rangle=0
\end{equation}

Thus Eq(\ref{Om}) can be rewritten as
\begin{equation}\label{Omiga}
\Omega=\sum_i\Omega_{i}^{\mathcal{C}},
\end{equation}
and the single-site matrice is given as
\begin{equation}
\Omega_{i}^{\mathcal{C}}=\sum_{i}\langle T^R_i\bar{g}^R C \bar{g}^AT^A_i\rangle.
\end{equation}
Substituting Eq.(\ref{Ti}) into Eq.(\ref{Omiga}), the terms with single
$t_i^{\mathcal{C}}$ vanish in the SSA, and we can finally obtain

\begin{equation}
\Omega_{i}^{\mathcal{C}}=\langle t_{i}^{\mathcal{C},R}\bar{g}^R C \bar{g}^At_{i}^{\mathcal{C},A}\rangle+\sum_{j\ne i}\langle t_{i}^{\mathcal{C},R}\bar{g}^R\Omega_{j}^{\mathcal{C}}\bar{g}^At_{i}^{\mathcal{C},A}\rangle.
\end{equation}
To solve above equation for $\Omega_{i}^{\mathcal{C}}$, we introduce, in coupling space
\begin{equation}
\sum_{j\ne i}\Omega_{j}^{\mathcal{C}}=\Omega-\Omega_{i}^{\mathcal{C}},
\end{equation}
$\Omega_{i,\mathcal{C}}$ can thus be rewritten as
\begin{widetext}
\begin{equation}
\begin{aligned}
\Omega_{i}^{\mathcal{C}}=\langle t_{i}^{\mathcal{C},R}[\bar{g}^R C \bar{g}^A]_i^{\mathcal{C}}t_{i}^{\mathcal{C},A}\rangle 
-\langle t_{i}^{\mathcal{C},R}[\bar{g}^R]_i^{\mathcal{C}}\Omega_{i}^{\mathcal{C}}[\bar{g}^A]_i^{\mathcal{C}}t_{i}^{\mathcal{C},A}\rangle+\langle t_{i}^{\mathcal{C},R}[\bar{g}^R\Omega\bar{g}^A]_i^{\mathcal{C}}t_{i}^{\mathcal{C},A}\rangle.
\label{Omegai}
\end{aligned}
\end{equation}
\end{widetext}
and $\langle t_{i}^{\mathcal{C},R}M_i^{\mathcal{C}}t_{i}^{\mathcal{C},A}\rangle=\sum_{Q,n} c_i^Q p_i^{n}t_i^{\mathcal{C},Q,n,R}M_i^{\mathcal{C}}t_i^{\mathcal{C},Q,n,A}$. Here, we can calculate the quantities $[\bar{g}^R C \bar{g}^A]_i^{\mathcal{C}}$, and $[\bar{g}^R\Omega\bar{g}^A]_i^{\mathcal{C}}$ by the integration
\begin{widetext}
\begin{equation}
\begin{aligned}
&[\bar{g}^R C \bar{g}^A]^{\mathcal{C}}_{i, {\vec{B}_i+\vec{T}_{\|}, \vec{B}_j+\vec{T}_{\|}^{\prime}}}
=\frac{1}{\mathcal{V}} \int d \vec{k} [\bar{g}^R C \bar{g}^A]_{\vec{B}_i, \vec{B}_j}(\vec{k}_{\|}) e^{-i \vec{k}_{\|}\cdot(\vec{T}_{\|}^{\prime}-\vec{T}_{\|})}
\end{aligned}
\end{equation}
\begin{equation}
\begin{aligned}
&[\bar{g}^R\Omega\bar{g}^A]^{\mathcal{C}}_{i,\vec{B}_i+\vec{T}_{\|}, \vec{B}_j+\vec{T}_{\|}^{\prime}}
=\frac{1}{\mathcal{V}} \int d \vec{k} [\bar{g}^R\Omega\bar{g}^A]_{\vec{B}_i, \vec{B}_j}(\vec{k}_{\|}) e^{-i \vec{k}_{\|}\cdot(\vec{T}_{\|}^{\prime}-\vec{T}_{\|})}.
\end{aligned}
\end{equation}
\end {widetext}

 Then, the quantity $\Omega_{i,\mathcal{C}}$ can be solved self-consistently from Eq.\ref{Omegai}. 

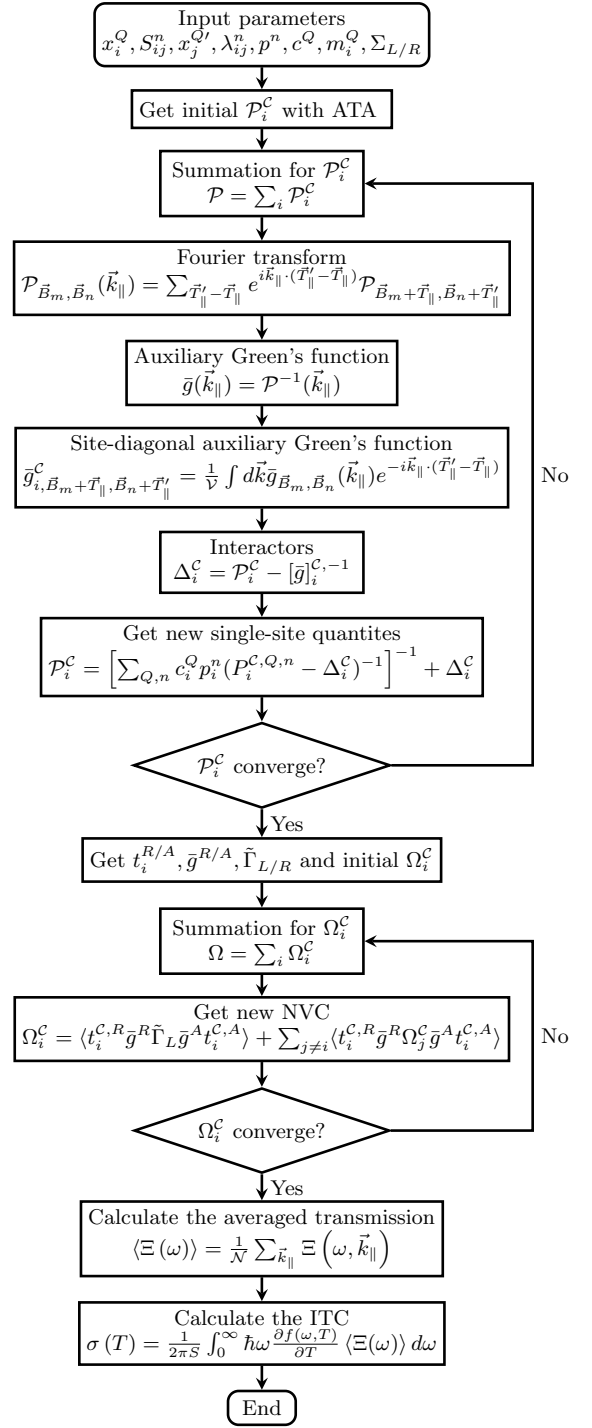
\begin{figure}[t]
  \centering
  \begin{tikzpicture}[node distance=1.4cm, scale=0.9, transform shape]
    \node (start) [startstop, align=center] {Input parameters \\$x_{i}^{Q},S_{ij}^{n},x_{j}^{Q\prime},\lambda _{ij}^{n},p^n,c^Q,m_{i}^{Q},\Sigma _{L/R}$};
    \node (pro1) [process, below of=start,node distance=1.1cm, align=center] {Get initial  $\mathcal{P} _{i}^{\mathcal{C}}$ with ATA };
    \node (pro2) [process, below of=pro1,node distance=1.1cm,align=center] {Summation for  $\mathcal{P} _i^\mathcal{C}$  \\$\mathcal{P} =\sum_i{\mathcal{P} _{i}^{\mathcal{C}}}$};
    \node (pro3) [process, below of=pro2, align=center] {Fourier transform \\$\mathcal{P} _{\vec{B}_m,\vec{B}_n}(\vec{k}_{\parallel})=\sum_{\vec{T}_{\parallel}^{\prime}-\vec{T}_{\parallel}}{e^{i\vec{k}_{\parallel}\cdot (\vec{T}_{\parallel}^{\prime}-\vec{T}_{\parallel})}}\mathcal{P} _{\vec{B}_m+\vec{T}_{\parallel},\vec{B}_n+\vec{T}_{\parallel}^{\prime}}$};
     \node (pro4) [process, below of=pro3,node distance=1.4cm, align=center] {Auxiliary Green’s function \\$\bar{g}(\vec{k}_{\parallel})=\mathcal{P} ^{-1}(\vec{k}_{\parallel})$};
     \node (pro5) [process, below of=pro4, align=center] {Site-diagonal auxiliary Green’s function \\ $\bar{g}_{i,\vec{B}_m+\vec{T}_{\parallel},\vec{B}_n+\vec{T}_{\parallel}^{\prime}}^\mathcal{C}=\frac{1}{\mathcal{V}}\int{d}\vec{k}\bar{g}_{\vec{B}_m,\vec{B}_n}(\vec{k}_{\parallel})e^{-i\vec{k}_{\parallel}\cdot (\vec{T}_{\parallel}^{\prime}-\vec{T}_{\parallel})}$};
    \node (pro6) [process, below of=pro5,node distance=1.4cm, align=center] {Interactors\\$\Delta _{i}^{\mathcal{C}}=\mathcal{P} _{i}^{\mathcal{C}}-\left[ \bar{g} \right] _{i}^{\mathcal{C},{-1}}$};
    \node (pro7) [process, below of=pro6,node distance=1.4cm, align=center] {Get new single-site quantites \\ $\mathcal{P} _{i}^{\mathcal{C}}=\left[ \sum_{Q,n}{c_{i}^{Q}p_{i}^{n}}(P_{i}^{
\mathcal{C},Q,n}-\Delta _{i}^{\mathcal{C}})^{-1} \right] ^{-1}+\Delta _{i}^{\mathcal{C}}$};
    \node (dec1) [decision, below of=pro7,node distance=1.6cm] {$\mathcal{P} _{i}^{\mathcal{C}}$ converge?};
    \node (pro8) [process, below of=dec1,node distance=1.4cm, align=center] {Get $t_{i}^{R/A},\bar{g}^{R/A},\tilde{\Gamma}_{L/R}$ and  initial $\Omega_i^{\mathcal{C}}$};
    \node (pro9) [process, below of=pro8,node distance=1.2cm, align=center] {Summation for $\Omega _{i}^{\mathcal{C}}$\\ $\Omega =\sum_i{\Omega _{i}^{\mathcal{C}}}$};
    \node (pro10) [process, below of=pro9,node distance=1.3cm, align=center] {Get new NVC\\$\Omega_{i}^{\mathcal{C}}=\langle t_{i}^{\mathcal{C},R}\bar{g}^R \tilde{\Gamma}_L \bar{g}^At_{i}^{\mathcal{C},A}\rangle+\sum_{j\ne i}\langle t_{i}^{\mathcal{C},R}\bar{g}^R\Omega_{j}^{\mathcal{C}}\bar{g}^At_{i}^{\mathcal{C},A}\rangle$};
    \node (dec2) [decision, below of=pro10,node distance=1.5cm] {$\Omega_{i}^{\mathcal{C}}$ converge?};
\node (pro11) [process, below of=dec2,node distance=1.6cm, align=center] {Calculate the averaged transmission \\  $\left< \Xi \left( \omega \right) \right> =\frac{1}{\mathcal{N}}\sum_{\vec{k}_{\parallel}}{\Xi \left( \omega ,\vec{k}_{\parallel} \right)}$};
\node (pro12) [process, below of=pro11,node distance=1.4cm, align=center] {Calculate the ITC \\ $\sigma \left( T \right) =\frac{1}{2\pi S}\int_0^{\infty}{\hbar}\omega \frac{\partial f(\omega ,T)}{\partial T}\left< \Xi (\omega ) \right> d\omega $};
    \node (stop) [startstop, below of=pro12,node distance=1.1cm] {End};

    \draw [arrow] (start) -- (pro1);
    \draw [arrow] (pro1) -- (pro2);
    \draw [arrow] (pro2) -- (pro3);
    \draw [arrow] (pro3) -- (pro4);
    \draw [arrow] (pro4) -- (pro5);
    \draw [arrow] (pro5) -- (pro6);
    \draw [arrow] (pro6) -- (pro7);
    \draw [arrow] (pro7) -- (dec1);
    \draw [arrow] (dec1) -- ++(4,0) -- node [right]{No} ++(0,8.6)--(pro2);
    \draw [arrow] (dec1) -- node [right]{Yes}(pro8);
    \draw [arrow] (pro8) -- (pro9);
    \draw [arrow] (pro9) -- (pro10);
    \draw [arrow] (pro10) -- (dec2); 
    \draw [arrow] (dec2) -- ++(4,0) -- node [right]{No} ++(0,2.8)--(pro9);
    \draw [arrow] (dec2) -- node [right]{Yes}(pro11) ;
    \draw [arrow] (pro11) -- (pro12); 
    \draw [arrow] (pro12) --(stop);
    
  \end{tikzpicture}
  \caption{Sketch of the ACPA-NVC self-consistent steps for quantum transport simulation of disordered nano-structures. }
  \label{loop}
\end{figure}

Then, the averaged transmission function can be obtained by, with $C=\tilde{\Gamma}_L$,
\begin{equation}
\begin{aligned}
\langle \Xi(\omega)\rangle=\bar{g}^R\tilde{\Gamma}_L\bar{g}^A\tilde{\Gamma}_R+\bar{g}^R\Omega\bar{g}^A\tilde{\Gamma}_R
\label{Xi1}
\end{aligned}
\end{equation}
with the $\vec{k}_{\|}$-resolved transmission function is given by 
\begin{equation}
\begin{aligned}
\langle \Xi(\omega,\vec{k}_{\|})\rangle=&\bar{g}^R(\vec{k}_{\|})\tilde{\Gamma}_L(\vec{k}_{\|})\bar{g}^A(\vec{k}_{\|})\tilde{\Gamma}_R(\vec{k}_{\|}) \\
&+\bar{g}^R(\vec{k}_{\|})\Omega(\vec{k}_{\|})\bar{g}^A(\vec{k}_{\|})\tilde{\Gamma}_R(\vec{k}_{\|}).
\label{Xi2}
\end{aligned}
\end{equation}
Here, in Eq.(\ref{Xi1}) and Eq.(\ref{Xi2}), the first term is called the coherent part, and the second is call the diffusive part describing the momentum relaxation due to multiple disorder scatterings. Thus, with the above ACPA-NVC method, we can calculate the ITC through the disordered devices with both the diagonal and off-diagonal disorders. Finally, we summarize the numerical implementation and simulation of quantum transport through disordered device structures in Fig.\ref{loop} as follows:

 (1)Input the parameters $x_{i}^{Q}$,$ S_{ij}^{n}$, $x_{j}^{Q\prime}$, $\lambda _{ij}^{n}$, $p^n$, $c^Q$, $m_{i}^{Q}$ for representing the (disordered) masses and force-constants in the central region and leads, and calculate the lead self-energies $\tilde{\Sigma}_{L/R}$;

(1)Get initial guess for $\mathcal{P}_{i}^{\mathcal{C}}$ (e.g.with averaged T-matrix approximation(ATA));

(2)Sum up $\mathcal{P}_{i}^{\mathcal{C}}$ to obtain the effective medium auxiliary Hamiltonian elements $\mathcal{P}$;

(3)Apply the Fourier transformation Eq.(\ref{FT}) to obtain $\mathcal{P}_{\vec{B}_m,\vec{B}_n}( \vec{k}_{\|})$;

(4) Calculate the auxiliary Green's function $\bar{g}(\vec{k}_{\|})$ by Eq.(\ref{gk});

(5) Calculate $[\bar{g}]_{i}^{\mathcal{C}}$ by Eq.(\ref{gii}) with the inverse Fourier transformation;

(6) Calculate the interactor $\Delta_i^{\mathcal{C}}$ by Eq.(\ref{delta});

(7) Calculate new single-site quantities $\mathcal{P}_{i}^{\mathcal{C}}$ by Eq.(\ref{Pi}) with the $\Delta_i^{\mathcal{C}}$;

(8) Repeat the steps $(2)-(7)$ until $\mathcal{P}_{i}^{\mathcal{C}}$ is converged;

(9) Calculate the averaged auxiliary Green's function $[\bar{g}]_i^{\mathcal{C},R/A}$,$[\bar{g}\tilde{\Gamma}_L\bar{g}]_i^{\mathcal{C},R/A}$, calculate $t_i^{\mathcal{C},Q,n,R/A}$ by Eq.(\ref{ti}) and give an initial guess $\Omega_{i}^{\mathcal{C}}$;

(10) Sum ${\Omega}_{i}^{\mathcal{C}}$ to obtain $\Omega$ and then calculate $[\bar{g}\Omega\bar{g}]_{i}^{\mathcal{C}}$;

(11) Obtain new $\Omega_{i}^{\mathcal{C}}$ by Eq.(\ref{Omegai});

(12) Repeat the steps $(10)-(11)$ until $\Omega_{i}^{\mathcal{C}}$ is converged;

(13) Calculate the averaged transmission function by Eq.(\ref{Xi1})and(\ref{Xi2});

(14) Calculate the ITC by Eq.(\ref{sigmaT}).

\section{NUMERICAL RESULTS AND DISCUSSIONS} \label{results}
In this section, we investigate the phonon transport through disordered $Ni|Ni_xPt_{1-x}|Pt$ interface with intermixed $Ni_xPt_{1-x}$ disordered layers as shown in Fig.\ref{IS} by applying the self-consistent ACPA-NVC method. The force constant used in our ACPA-NVC calculations, including the averaged-value FCD and Anderson-type FCD, for different $x$ is calculated by the Vienna ab initio simulation package(VASP)\cite{VASP1,VASP2,Potential1,Potential2,Potential3,Potential4,Potential5,Potential6,SQS} with density functional perturbation theory(DFPT)\cite{DFPT}. As shown in Ref.\onlinecite{ACPA4}, $Ni_xPt_{1-x}$ with high concentration features important distribution in FCD, namely important AFCD effects. In practical  ACPA-NVC calculations, to ensure the convergence of transmission function $\Xi(\omega)$, $100\times100$ transverse $k_{\|}$ mesh is used to converge the 2D-BZ integration and  100 frequency points are calculated to the converge the ITC $\sigma$. 

\subsection{Validation for the ACPA-NVC method} 
\begin{figure}[htbp]
			\centering

\includegraphics[width=0.45\textwidth]{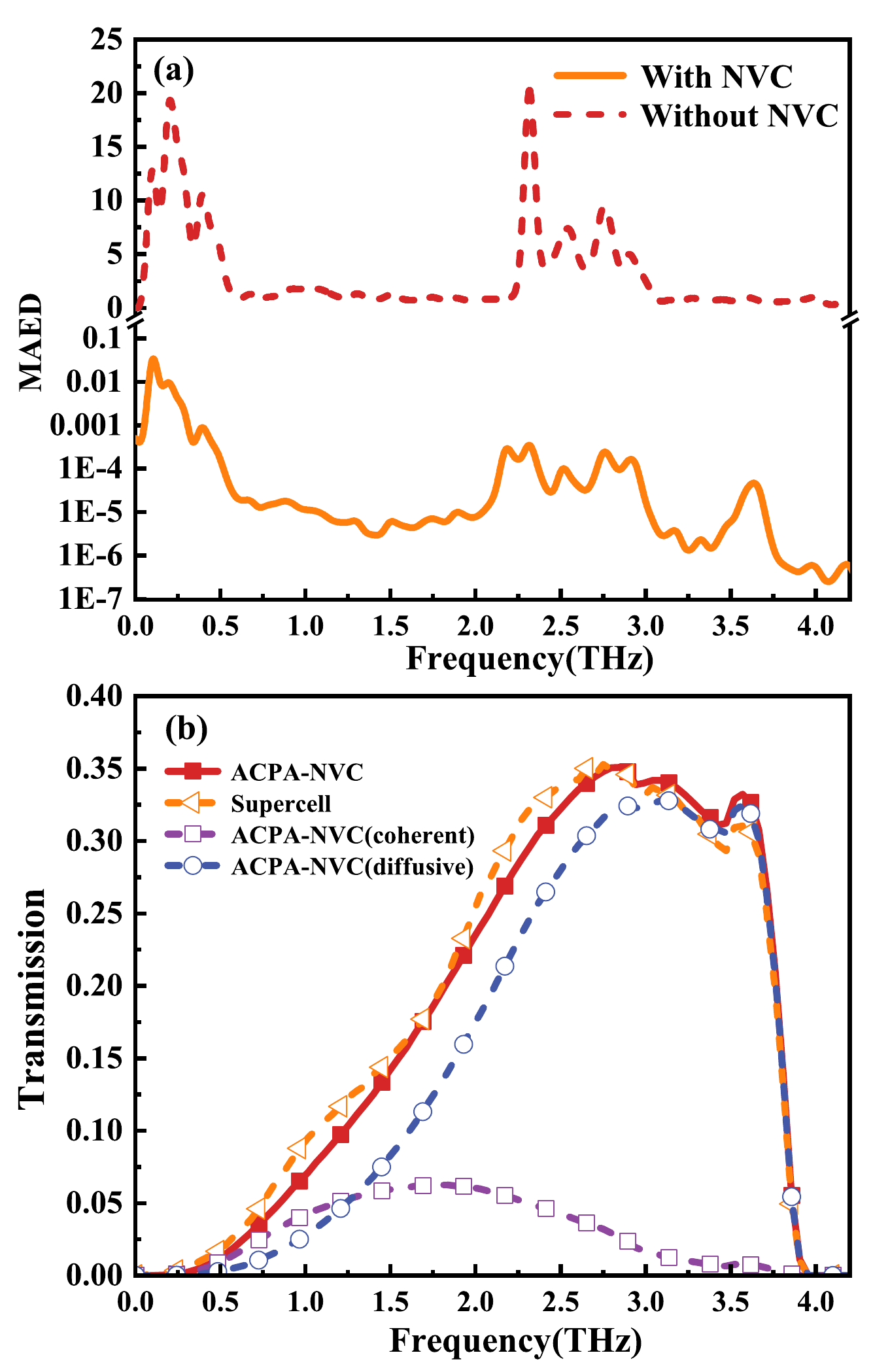}
	\caption{Validation of the ACPA-NVC formalism with the disordered junction $Ni|(4ML)Ni_{0.5}Pt_{0.5}|Pt$: (a)the maximum absolute element of the difference (MAED) (versus frequency) between the two sides of the Eq.(\ref{FDTeq1}). The MAED results with and without NVC are presented. (b) the transmission versus frequency for supercell method, ACPA-NVC method, coherent part and diffusive part of ACPA-NVC method.}	
\label{fig4}		    		    
\end{figure}

It is known that at the equilibrium, namely $f_L = f_R$, the lesser Green's function of the device satisfies the fluctuation-dissipation theorem, namely 
\begin{equation}
 \bar{g}^{<}= \bar{g}^{A}-\bar{g}^{R}.\label{FDTeq1}
\end{equation}
 As an important test of  the present implimentation of ACPA-NVC , we have checked the fluctuation-dissipation theorem, in which the direct calculation of $\bar{g}^{<}$ requires the NVC for each $k_{\|}$as follows,
\begin{equation}
\bar{g}^{<}(k_{\|})=\bar{g}^R\left(k_{\|}\right) \tilde{\Sigma}^{<}\left(k_{\|}\right) \bar{g}^A\left(k_{\|}\right)+\bar{g}^R\left(k_{\|}\right) \Omega\left(k_{\|}\right) \bar{g}^A\left(k_{\|}\right),\label{FDTeq2}
\end{equation}
while $\bar{g}^{<}(k_{\|})=\bar{g}^A(k_{\|})-\bar{g}^R(k_{\|})$  is starightforward with ACPA. Here the lesser self-energy of leads $\tilde{\Sigma}^{<} = if_l\tilde{\Gamma}_L+ if_R \tilde{\Gamma}_R$.
In this test, we consider the junction structure $Ni|4(ML)Ni_{0.5}Pt_{0.5}|Pt$ with a disordered interface containing 4 atomic monolayers (ML) of intermixed  $Ni_{0.5}Pt_{0.5}$.
The quantities in Eq.(\ref{FDTeq1}) are all matrices, and to visualize that both sides of Eq.(\ref{FDTeq1}) are equal, we take the maximum absolute elements of the difference(MAED) between the two sides of the equation for all $k_{\|}$ in the 2D BZ with $100\times100$ mesh. To show the necessity of NVC, we also show the MAED without the NVC part, namely taking $\Omega=0$ in Eq.(\ref{FDTeq2}). As shown in Fig.\ref{fig4}, the large magnitude of MAED without NVC (in red) present a large error for the relation of fluctuation-dissipation theorem, while the inclusion of NVC term greatly reduces the error to be negligible in the results of MAED (in yellow). It is clear that, compared to the results without NVC, including the NVC presents more than three orders of magnitude reduction at different frequencies. Satisfing the fluctuation-dissipation theorem provides an important test for the ACPA-NVC method and our correct implementation of both ACPA and NVC. 

%
%
%

As a test on the phonon transport calculation, in Fig.\ref{fig4}(b),  we investigate the phonon transmission (versus frequency $w$) for mass disorder(MD)+AFCD calculated by ACPA-NVC, and compare with the supercell calculation. We calculate phonon transmission through the disordered structure $Ni|(4ML)Ni_{0.5}Pt_{0.5}|Pt$.
For supercell calculations, we use the transverse supercell size of {$4\times4$} atoms with  $30\times30$ $k_{\|}$-sampling in BZ and 500 random atomic configurations for disorder average. It is clear, for the whole frequency range as shown in Fig.\ref{fig4}(b), the ACPA-NVC result (in red square) presents quite good agreement with the supercell calculations (in yellow empty triangle), showing the important effectiveness of ACPA-NVC for capturing the multiple disorder scatterings in the present system. For example, at $w=1.69$THz, the ACPA-NVC transmission is 0.175, agreeing well with the supercell result 0.177. As shown, the coherent part of ACPA-NVC  (in empty square) dominates the transmission 
for the very low frequency phonon (with $w \le 1.2$THz) which is immune to disorder scatterings. However, at the high frequency, the diffusive part, accounting for the multiple disorer scatterings, presents dominant contribution, while the coherent part shows a minor contribution, demonstrating the neccessity for developing the ACPA-NVC for treating (A)FCD or the general off-diagonal disorders in quantum transport. For example, at $w=2.89$THz, the coherent part is $0.023$, compared to the diffusive part $0.324$. At higher frequencies, the coherent part becomes negligibly small, showing the importance of the NVC term. The agreement between ACPA-NVC and supercell calculations illustrates the efffectiveness of ACPA-NVC for quantum tansport simulation of realistic disordered structures with off-diagonal disorders.

\subsection{Phonon transport through disordered Ni/Pt interface}\label{concentration}
\begin{figure*}[htbp]
\centering
\includegraphics[width=0.9\textwidth]{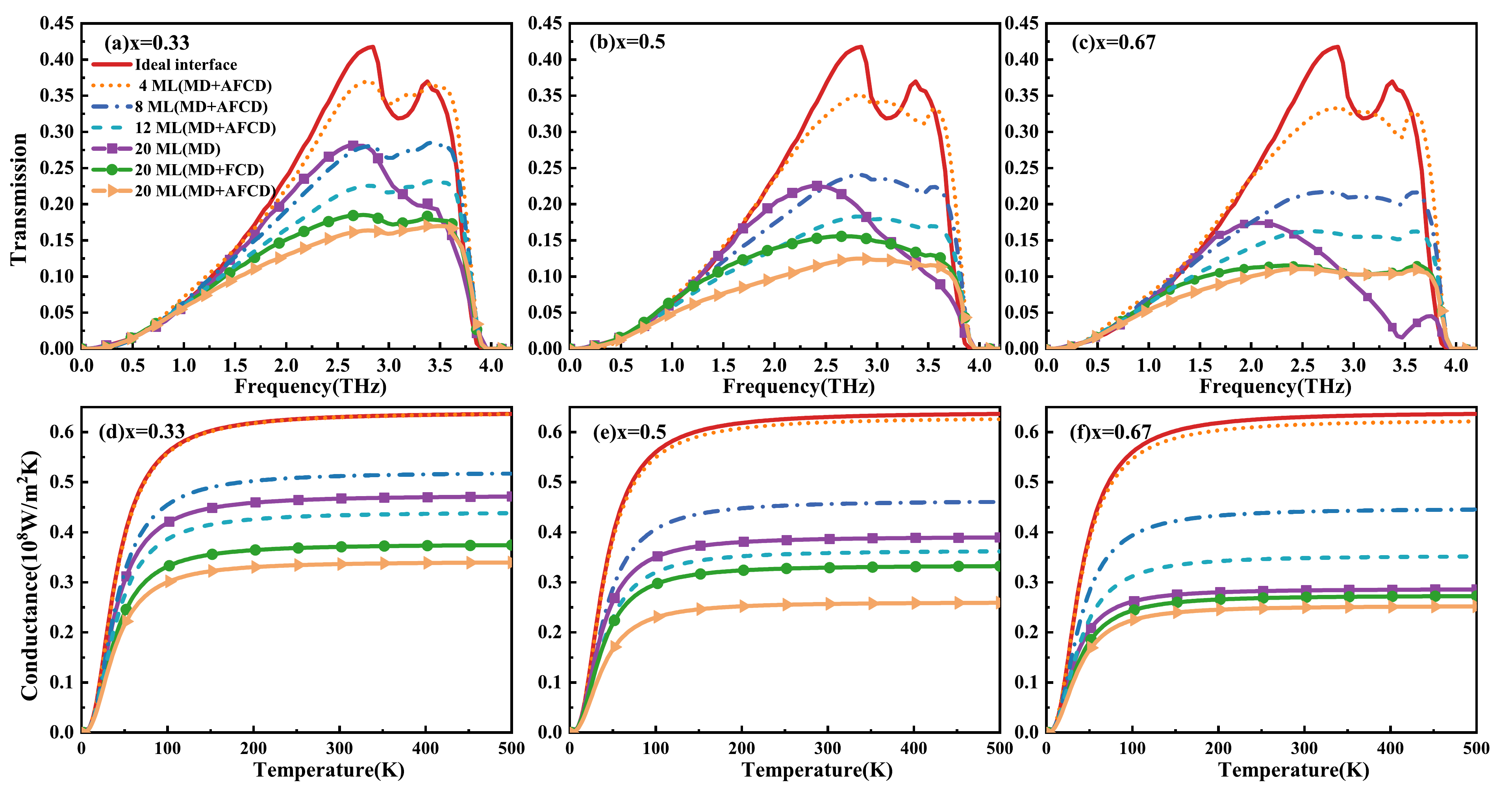}
    \caption{The transport properties for the disordered junctions  $Ni|Ni_{x}Pt_{1-x}|Pt$ with different disorder types and different length of disordered atomic layers. (a-c) shows the transmission versus frequency for the respective $x=0.33$,$0.5$ and $0.67$; (e-g) presents the interface thermal conductance versus temperature for the respective $x=0.33$,$0.5$ and $0.67$.}
\label{fig5}
\end{figure*}

In this section, we investigate the important influence of disorder scattering on phonon transport through disordered Ni/Pt interface with the ACPA-NVC method. We consider the device structure $Ni|Ni_{x}Pt_{1-x}|Pt$ containing different number of disordered atomic monolayers and all the force constants are abtained from VASP by calculating 108-atom supercells for different concentrations (for each $x$ we use 3 different atomic configurations) and the AFCD decomposition parameters are obtained with the method presented in Ref.\onlinecite{ACPA4}. Fig.\ref{fig5}(a)-(c) shows the transmission versus frequency for 20 ML $Ni_xPt_{1-x}$  with MD+AFCD (in yellow) for the respective $x=0.33,0.5,0.67$, and compare with the results of MD (in purple), and MD+FCD (in green,in which the averaged value of force constants are used). In Fig.\ref{fig5}(a)-(c), we see that, for the low frequencies $w\le1.2$THz, the transmission remains almost unchanged with varying disorder concentration x and changing the disorder types, presenting the immunity of long-wave-length phonons to disorders at the nano-scale. However, for each $x$, significant deviation in transmission between different disorder type appears for $w\ge1.3$THz, presenting the significantly enhanced disorder scattering to high-frequency phonons. For example, for x=0.5, at $w=2.41$THz, compared to the MD value of 0.225, MD+FCD decrease the transmission to 0.153, while MD+AFCD further reduces it to 0.115, presenting the importance for considering the FCD and its Anderson-type distribution. However,
around the cutoff frequency $w=3.91$THz, including FCD or AFCD can significantly increase the transmission compared to the result of MD, for example,for $x=0.5$, at $w=3.62$THz, the MD+FCD and MD+AFCD transmission results are the respective 0.126 and 0.113, larger than the MD value 0.089, presenting the incorporative effect of MD and (A)FCD.~\cite{ZhaiJX2019, ZhaiJX} As shown in Fig.\ref{fig5}(a)-(c) for all concentrations, the inclusion of the distribution of force constants in MD+AFCD generally reduces the transmission compared to the MD+FCD calculations for all high frequencies, which is consistent with the broadened linewidth function reported in Ref.\onlinecite{ACPA4}. It is clear that, as increasing $x$ from 0.33 to 0.67, the  transmission is decreased for different types of disorders. Fig.\ref{fig5}(a)-(c) also presents the transmission results of MD+AFCD for different length of disordered layers, including 0ML (perfect interface), 4ML,8ML and 12ML, to compare with the results of 20ML$Ni_xPt_{1-x}$. It is clear that the increase of disordered layers can significantly reduce the transmission of high-frequecy phonons, while the reduction in low-frequency transmission is rather small. 

\begin{figure}[htbp]
    \centering
    \includegraphics[width=0.45\textwidth]{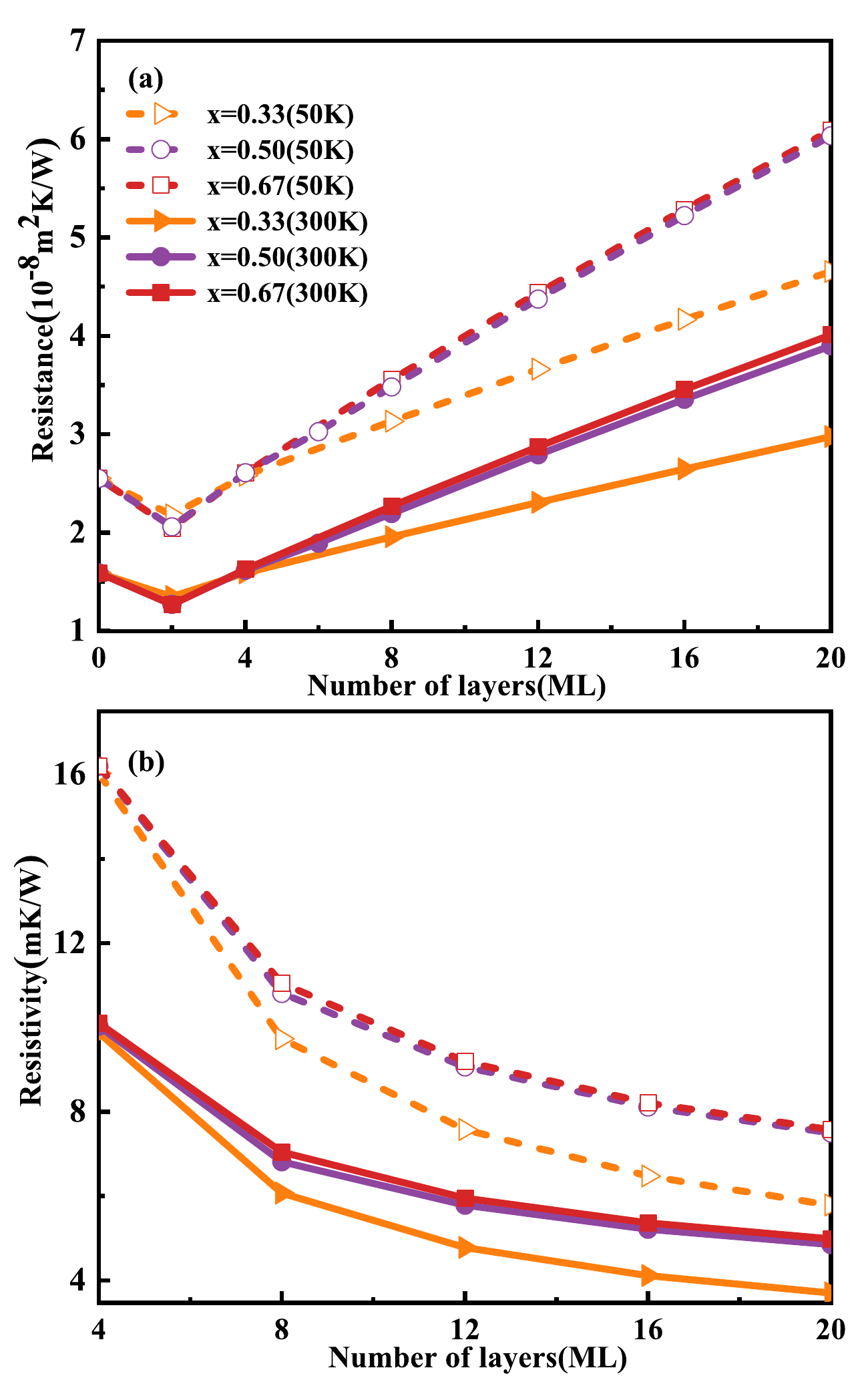}
    \caption{(a) The resistance versus number of disordered layers at 50K and 300K; (b) thermal resistivity versus number of disordered layers at 50K and 300K.}
\label{fig6}
\end{figure}
Fig.\ref{fig5}(d)-(f) shows the interface thermal conductance(ITC) versus temperature for the respective $x=0.33,0.5,0.67$ with different types of disorders and different number of disordered layers. 
It is clear that, as the temperature increases, all the ITC grow very rapidly before 100K and gradually approaches saturation around 300K. For different $x$ as shown, compared to the MD result for the 20ML, MD+FCD reduces the ITC and MD+AFCD further reduces the result.In particular, for $x=0.5$ at 300K, it is clear that MD+FCD and MD+AFCD calculations present the respective reduction of $14.8\%$ and $33.4\%$ in magnitude compared to the MD value, presenting the importance of including the FCD and its Anderson distribution.  Compred to the $x=0.33$ and 0.5, the reduction in ITC with FCD and AFCD for $x=0.67$ is relatively smaller, e.g. at 300K, the ITC values are 0.27$\times 10^8$W/m$^2$K, 0.25$\times 10^8$W/m$^2$K for the resepective MD+FCD and MD+AFCD, compared to the MD value $0.28$$\times 10^8$W/m$^2$K, attributing to the enhancement of transmission by the corparative effects of MD and FCD(and AFCD) at the high frequency. As a result the MD+AFCD results for x=0.5 and 0.67 are quite close in magnitude, e.g. at 300K, the ITC values are the respective $0.26$$\times 10^8$W/m$^2$K and $0.25$$\times 10^8$W/m$^2$K. However, for the MD and MD+FCD results, as increasing $x$ from 0.33 to 0.67, ITC presents the evident reduction. For example, at 300K, for the MD+FCD, the ITCs are $0.33$$\times 10^8$W/m$^2$K and $0.27$$\times 10^8$W/m$^2$K for $x=0.5$ and 0.67, compared to 0.37$\times 10^8$W/m$^2$K for $x=0.33$. Fig.\ref{fig5}(d)-(f) also presents the ITC for different length of disordered layers (including perfect, 4ML, 8ML and 12ML). ITC is quickly decreased by increasing the disordered layers, presenting increased disorder scattering on the high-frequency phonons.

In Fig.\ref{fig6} (a), we present the  interface thermal resistance(ITR) versus the number of disordered $Ni_xPt_{1-x}$ layers ( from $l=0$ML to 20ML) for the MD+AFCD for different $x$ at T=50K and 300K. 
The ITR is defined as the inverse of ITC, namely
\begin{equation}
R=1/\sigma.
\end{equation}
The ITR contains both the thermal resistance contributed by the interface scattering and the resistance induced by the diffusive scattering in disordered layers. As shown in Fig.\ref{fig6}(a), It is noted that the ITR  for $l=2$ML is lower than the result of ideal interface ($l=0$ML). The introduction of short disordered layers alters the interface characteristics, thereby reducing scattering at the interface. For both temperatures,  we observe that all the ITR results increase rather linearly with the length of the disordered layers in a short length scale.   It is clear that the ITR results of  $x=0.5$ and $x=0.67$ are quite close to each other, while the ITC of $x=0.33$ is importantly lower than other results. For example, for $l=20$ML at 300K, the ITCs of $x=0.5$ and $x=0.67$ are the repective values of 3.9$\times10^{-8}$m$^2$K/W, 4.0$\times10^{-8}$m$^2$K/W, while the result of $x=0.33$ is 3.0$\times10^{-8}$m$^2$K/W. As also shown in Fig.\ref{fig6}(a), compared to $T=300$K, lowering the temperature to $50$K significantly enhances the ITR, for example, for $l=16$ ML, ITR values are 3.4$\times10^{-8}$m$^2$K/W and 2.6$\times10^{-8}$m$^2$K/W at T=300K for $x=0.5$ and $0.33$, significantly lower than the corresponding values 5.2$\times10^{-8}$m$^2$K/W, 4.2$\times10^{-8}$m$^2$K/W at $50$K.

%

The thermal resistivity is defined as $\rho=R/l$. As $l$ approaches macroscopic scale, $\rho $ approaches the bulk limit of resistivity, that is,  namely the thermal resistivity of the bulk $Ni_xPt_{1-x}$.
Fig.\ref{fig6}(b) shows the thermal resistivity versus the number of disordered layers (for $l\le20$ ML) under the different temperatures. At both 50K and 300K, the thermal resistivity decreases with an increasing number of layers for all concentrations $x$. $x=0.67$ exhibits the highest resistivity, and $x=0.33$ presents the lowest. for example, at $l=20$ML, by increasing $x$ from 0.33 to 0.67, the thermal resistivity  is increased from 5.78mK/W to 7.57mK/W at 50K, and from 3.70mK/W to 4.98mK/W at 300K, presenting the important effects of disorder scattering. The development of ACPA-NVC method can provide an effective approach for simulating the important scattering effects of both diagonal and off-diagonal disorders on thermal transport, which is important for disordered material and structure design for thermal management.

\section{CONCLUSION}\label{conclusion}
In summary, we have developed and implemented the general ACPA-NVC method to treat both the diagonal and off-diagonal disorders in quantum transport simulation.  ACPA-NVC provides an effective mean-field approach for treating both diagonal and off-diagonal disorders. We have demonstrated that accounting for force-constant disorder is crucial for accurate modeling of the thermal transport properties of nanoscale devices. By calculating the disordered Ni/Pt interface, we show the important effects of mass and force-constant disorders on the thermal transport through disordered interface, and highlight the important influence of Anderson-type force-constant disorder on the interface thermal transmission. It is straightforward to generalize the ACPA-NVC approach presented herein to simulate a wide range of disordered nanoelectronic systems, providing a effective approach to reveal the quantum transport properties of disordered systems.
\section{ACKNOWLEDGMENTS}

Y.K. acknowledges financial support from NSFC (Grant
No. 12227901). The authors thank the HPC platform of
ShanghaiTech University for providing the computational facility.

\bibliographystyle{apsrev4-2.bst}

\bibliography{Trans}

\end{document}